\documentclass[doublecol]{epl2} 

\usepackage{amsmath}

\newcommand{\average}[1]{\ensuremath{\langle#1\rangle} }

\title{Evolution of favoritism and group fairness in a co-evolving three-person ultimatum game}
\shorttitle{Co-evolving three-person ultimatum games} 

\author{Hirofumi Takesue\inst{1} \and Akira Ozawa\inst{2} \and So Morikawa\inst{2}}
\shortauthor{H. Takesue \etal}

\institute{ 
\inst{1} Graduate Schools for Law and Politics, University of Tokyo - 7-3-1, Bunkyo, Hongo, Tokyo, 1130033, Japan\\
\inst{2} Graduate School of Engineering, University of Tokyo - 7-3-1, Bunkyo, Hongo, Tokyo, 1138656, Japan
}

\abstract{
The evolution of fairness in dyadic relationships has been studied using ultimatum games. However, human fairness is not limited to two-person situations and universal egalitarianism among group members is widely observed. In this study, we investigated the evolution of favoritism and group fairness in a three-person ultimatum game (TUG) under a co-evolutionary framework with both strategy updating and partner switching dynamics. In the TUG, one proposer makes an offer to two responders and the proposal is accepted at the group level if at least one individual responder accepts the offer. Investigating fairness beyond dyadic relationships allows the possibility of favoritism because the proposer can secure acceptance at the group level by discriminating in favor of one responder. Our simulation showed that the proposer favors one responder with a similar type when the frequency of partner switching is low. In contrast, group fairness is observed when the frequency of partner switching is high. The correlation between strategy and neighborhood size suggested that partner switching influences the strategy through the proposer's offer rather than through the responder's acceptance threshold. In addition, the average degree negatively impacts the emergence of fairness unless the frequency of partner switching is high. Furthermore, a higher frequency of partner switching can support the evolution of fairness when the maximum number of games in one time step is restricted to smaller values.}

\begin{document}

\maketitle

\section{Introduction}
The evolution of cooperation is one of the most actively investigated subjects in the physical and biological sciences \cite{Nowak2006, Szabo2007}. It is known that \textit{fair division} of the benefit of collaborative behavior is important for maintaining cooperative relationships \cite{Tomasello2009}. Thus, the fairness of resource division has been studied using ultimatum games \cite{Guth1982}. In ultimatum games, two players (a proposer and a responder) divide the resource. The proposer makes an offer regarding the resource division. If the responder accepts the offer, the resource is divided accordingly. If the responder rejects the offer, both players gain nothing.

The standard equilibrium notion in classical game theory predicts that the proposer almost monopolizes the resource. The responder gains nothing by rejection, so they should accept any positive offer. Expecting this reaction, a rational proposer should claim most of the resource. However, experimental evidence has repeatedly falsified this prediction \cite{Camerer2003}. An excessively low offer is often rejected, and the proposer offers nearly half of the resource to the responder. These observations are explained better by a model that incorporates the disutility due to inequity \cite{Fehr1999}.

Many theoretical models have been proposed to explain the evolutionary origin of fairness. One approach stresses that the opportunity to choose the interaction partner is crucial for the evolution of fairness \cite{Andre2011a}. Other studies have shown that error \cite{Santos2015a} and weak selection \cite{Rand2013} can also explain the preference for fairness in ultimatum games. The roles of reputation \cite{Nowak2000} and empathy \cite{Page2002} have also been investigated.

Among the mechanisms proposed for the evolution of fairness, the effect of the network (spatial) structure is among the most intensively studied. A seminal study has shown that introducing lattice structure can facilitate the emergence of fairness \cite{Page2000}. This positive effect of the network structure was also confirmed using more complex network structures \cite{Kuperman2008}. The role of the network structure has also been studied in combination with other mechanisms, including empathy \cite{Sinatra2009, Szolnoki2012}, the fineness of the strategy \cite{Szolnoki2012a}, role switching \cite{Wu2013}, allocation mechanism of divided resource \cite{Wang2014a,Chen2015a}, migration \cite{Wang2015d} and simple strategy updating after the breakdown of bargaining \cite{Duan2010}.

In addition, recent research considers the possibility that both the interaction structure and the strategy employed by players could co-evolve. In these studies, players can change interaction partners based on the neighbors' strategy or the payoff from the game. The results obtained in previous studies have shown that fairness can evolve more easily if opportunities exist for partner switching (choice) \cite{Deng2011, Gao2011, Miyaji2013, Yang2015}. The role of partner switching (choice) has also been investigated with respect to the evolution of cooperation. Previous studies have shown that partner switching (choice) enhances cooperation in the prisoner's dilemma \cite{Zimmermann2004, Pacheco2006a, Santos2006, Fu2008, Szolnoki2008b, VanSegbroeck2008, Fu2009, Szolnoki2009b, Tanimoto2009, Perc2010, Cong2014, Chen2016} and public goods game \cite{Wu2009a, Wu2009c, Zhang2011}.

These previous studies on the ultimatum game have deepened our understanding of fairness in \textit{dyadic} relationships. However, human fairness is not limited to two-person situations. Indeed, Boehm \cite{Boehm2012} noted that universal egalitarianism among group members is widely observed in hunter gatherers and tribal agriculturalists. Despite this empirical observation, there have been few theoretical analyses of group fairness. In contrast, the evolution of cooperation in public goods games on networks has been widely investigated \cite{Hauert2002, Santos2008, Szolnoki2009c, Helbing2010, Shi2010, Szolnoki2010a, Xu2010a, Szolnoki2011b, Chen2012, Wu2014a, Chen2015, Chen2016a}.

In this study, we investigated the evolution of \textit{group} fairness in the three-person ultimatum game (TUG) based on numerical simulations. In the TUG, one proposer makes a proposal regarding resource division to two responders. The proposal is accepted, and resource division occurs if \textit{at least} one responder accepts the offer. In the TUG, the proposer can secure the support of one responder while excluding the rest. As a result, the possibility of \textit{favoritism} is introduced. The possibility of favoritism questions the effectiveness of the mechanism that is assumed to support the evolution of group fairness. For example, Boehm \cite{Boehm2012} suggested that punishment via the coalition of weaker individuals can support the emergence of group fairness. However, a favored agent has little incentive to conduct punishment, so a coalition may be unstable.

Some important previous studies \cite{Santos2015, Santos2016} investigated the multiplayer ultimatum game in a well-mixed population. In the multiplayer ultimatum game, one proposer offers part of the resource to multiple responders and each responder accepts or rejects the offer. These studies have shown that the minimum number of individual acceptances needed for group level acceptance is important. Here, the proposer has to offer the \textit{same} amount of the resource to all responders, and the possibility of favoritism is not considered. In fact, one study \cite{Santos2016} noted that the effect of allowing the proposer to target offers to specific responders is an open question.

In this study, we introduced a co-evolutionary mechanism where the strategies used in the game and individual partnerships can both evolve. Our simulations showed that the ratio of the frequency of partner switching events relative to strategy updating events has a profound effect on the distribution of the resource in the group.

\section{Model}
First, we explained the TUG. In the TUG, there is one proposer and two responders. The proposer offers $p_A$ and $p_B$ to the two responders. Who will be offered $p_A \ (p_B)$ is determined by the agents' \textit{type}, $\theta$ (range: 0-1). The distance between two agents $i$ and $j$ is defined as $r_{ij} = \min(|\theta_i - \theta_j|, 1- |\theta_i - \theta_j|)$. Note that this definition means that the type is a circular variable where 0 and 1 are equivalent. This circular nature ensures that no agent is in advantageous position because of its type. The proposer offers $p_A \ (p_B)$ to the responder if their distance from the proposer is small (large). If $p_A > p_B (p_A < p_B)$, then the proposer favors (disfavors) the responder with a similar type. Without $\theta$, even if $p_A \neq p_B$, agents decide who will be favored randomly in each game and their behavior cannot be interpreted as favoritism in the long term. Each responder compares their minimum demand $q$ and $p_A \ (p_B)$ and accepts or rejects the offer. We introduced the group decision variable ($g$), which takes a value of 1 if the proposal is accepted at the group level and 0 otherwise.

The group makes a decision based on the proposal to offer $p_{iA} \ (p_{iB})$ to $j \ (k)$ in the following manner.
\begin{equation*}
g_{ijk} =
\begin{cases}
1, & \text{if}\ p_{iA} \geq q_j \lor p_{iB} \geq q_k \\
0, & \text{otherwise}
\end{cases}
\end{equation*}
Thus, the proposal is accepted if at least one responder accepts the offer. Using $g_{ijk}$, the payoff from the game for the proposer $i$ is calculated as $\pi_i = g_{ijk}[1 - (p_{iA} + p_{iB})]$. The payoff for the responders $j$ and $k$ are calculated as $\pi_j = g_{ijk}p_{iA}$ and $\pi_k = g_{ijk}p_{iB}$, respectively. Classical game theory predicts that the proposer will monopolize the resource in the same manner as in the two-person ultimatum game \cite{Santos2016}. Furthermore, the proposer obviously has a strong incentive to ignore one responder because their approval is not needed for group level acceptance.

Next, we explained the evolutionary process for the network structure and the strategy in the TUG. Let us assume that $N$ agents are located in the network, which is defined by the neighbors of each agent. The edges between agents represent social relationships. Initially, all agents have the same number of edges ($\average{k}$), which are randomly connected to other agents \cite{Santos2005a}. The size of the proposal has random values under the restriction that the sum of the proposals, $p_A + p_B$, does not exceed 1. The responder's minimum acceptance threshold also has a random value. The type ($\theta$) also follows a standard uniform distribution, $U(0, 1)$. In each time step, a strategy updating event or a partner switching event occurs. 

Strategy updating occurs with a probability of $1-w$. We used a link-based update rule to reduce the effects of large degree nodes in the neighborhood \cite{Fu2009}. This rule is similar to the ``pairwise comparison process'' in the well-mixed population model \cite{Traulsen2006}. In a strategy updating event, we first chose one edge ($E_{ij}$) randomly. Next, agent $i \ (j)$ plays TUG with their direct neighbors $k_i \ (k_j)$ times, where $k_i \ (k_j)$ is node $i$'s ($j$'s) number of neighbors. The number of games corresponds to the ordinary evolutionary game on the graph, where agents engage in two-person games with all their neighbors. In each game, two agents who play TUG with $i \ (j)$ are chosen randomly from the neighbors and the proposer is also chosen randomly from the three agents. Agents $i$ and $j$ gain a payoff from each game and accumulate total payoffs of $\Pi_i$ and $\Pi_j$, respectively. Next, $\Pi_i$ and $\Pi_j$ are compared and strategy updating occurs. Specifically, the strategy of agent $j$ replaces that of agent $i$ with a probability of
\begin{equation*}
P(s_i \leftarrow s_j) = [1 + \exp(-\beta(\Pi_j - \Pi_i))]^{-1}.
\end{equation*}
The value of $\beta$ is the intensity of selection ($\beta \to 0$ leads to random drift whereas $\beta \to \infty$ leads to imitation dynamics); otherwise, the strategy of agent $i$ replaces that of agent $j$. As a result, one of the two agents copies the other agent's strategy. In addition, the agent's type is copied at the same time. This corresponds to the assumption in previous studies regarding the evolution of favoritism where the strategies of agents as well as group membership evolve according to the payoff \cite{Fu2012}.

We assume that a small error accompanies the copying of attributes, where the error follows a uniform distribution, $U(-\epsilon, \epsilon)$. Note that if the strategic variable ($p_A$, $p_B$ or $q$) takes value outside the defined condition (range: 0-1), then we set the value to the nearest boundary value. In addition, if the sum of $p_A$ and $p_B$ exceeds 1, then we compressed it to one while keeping the ratio between $p_A$ and $p_B$. We also set the value of the type ($\theta$) to $\theta-1 (\theta+1)$ if it exceeds 1 (goes below 0) due to its circular nature.

A partner switching event is chosen with a probability of $w$. In a partner switching event, one agent ($i$) is chosen randomly and plays TUG with their direct neighbors $k_i$ times in the same manner as the strategy updating event. The agent decides the continuation of the social relationship based on the payoff from the games. Specifically, the agent finds one game where they gained the smallest payoff and breaks the relationship with the two neighbors who participated in that game \cite{Deng2011, Zhang2011}. If the agent earned the same smallest payoff in multiple games, one game is chosen randomly. Next, agent $i$ creates new links with two randomly chosen agents. We imposed the restriction that an agent with two edges does not lose links so that they can participate in TUG if they are a focal agent.

Intuitively, partner switching appears to engender a generous offer because responders will not be satisfied with a small offer and they will sever the link. In contrast, link adaptation appears to foster a ``rational'' acceptance threshold because rejection may cause the negotiation to break down (with the smallest payoff, 0). These opposite possible effects complicate the prediction.

\section{Results}
To investigate the emerging resource distribution in the TUG, we conducted numerical simulations. The simulations continued for $2 \times 10^7$ time steps, and we computed the quantities of interest by averaging over the values from the last $10^4$ time steps. We conducted 50 independent simulations for each combination of parameters and calculated the mean values from these simulation runs.

\begin{figure}[tbp]
\centering
\includegraphics[width = 55mm, trim= 0 15 0 0]{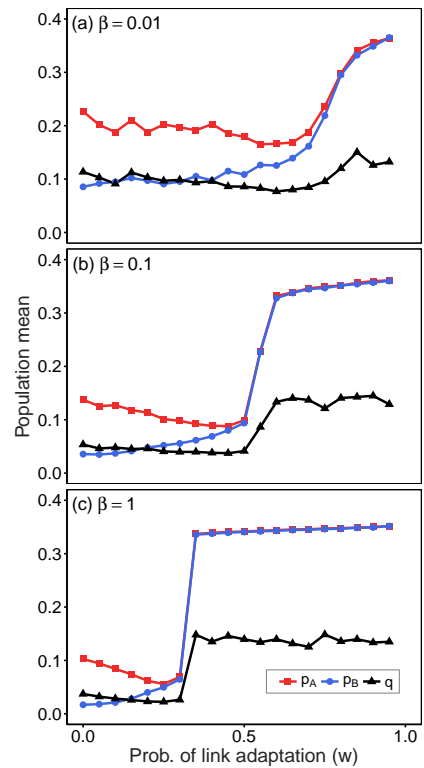}
\caption{\small Average offer ($p_A$, $p_B$) and acceptance threshold ($q$) as a function of $w$. The proposer favored one responder when $w$ was small. As $w$ increased, the proposer stopped discriminating and the offer size increased sharply when $w$ exceeded some specific value. The acceptance threshold also increased steeply together with the offer size. Parameters: (a) $\beta = 0.01$, (b) $\beta = 0.1$, (c) $\beta = 1$; fixed $N = 1000, \average{k} = 20$, and $\epsilon = 0.01$.}
\label{fig_papbq_w}
\end{figure}

First, we observed the impact of the frequency of partner switching on the TUG results. The parameter $w$ controls the frequency of partner switching, and a larger value implies that agents have more opportunities to adjust their social relationships. Figure \ref{fig_papbq_w} shows the mean values of $p_A$, $p_B$, and $q$ as a function of $w$ for different values of $\beta$. The figure shows that $p_A$ was larger than $p_B$ with a smaller value of $w$, which means that proposers consistently discriminated in favor of one responder with a similar type. Proposers treated one responder indifferently because this responder was strategically irrelevant for securing support at the group level, and the type functioned as a \textit{tag} when deciding the favored agent. The sum of the offer for two agents was larger when selection was weak (small $\beta$), and this result is similar to that obtained in previous studies \cite{Rand2013, Santos2015}. However, the condition that induces fairness in the two-person ultimatum game led to \textit{favoritism} rather than \textit{group} fairness.

Figure \ref{fig_papbq_w} also shows that as the value of $w$ increased, the value of $p_A$ started to decrease, whereas that of $p_B$ started to increase. Proposers stopped discriminating responders to prevent the relationship being severed by the indifferently treated responder. This increase in the offer raised the probability of $p_B$ being accepted, which deteriorated the importance of the favored agent in securing the $group$ level acceptance. Taking the result of $\beta = 0.1$ as an example, when $w = 0.05$, the number of cases where only $p_A$ was accepted was 13.7 times larger than the cases where only $p_B$ was accepted. Conversely, this ratio was only 1.8 when $w = 0.4$ and the importance of favored agents actually diminished. As a result, proposers could lower the degree of favoritism while keeping their own share ($p_A + p_B$).

In addition, the values of $p_A$ and $p_B$ both started to increase as $w$ increased further. In this situation, the proposers stopped monopolizing the resource and \textit{group} fairness was achieved. We noted that with a very large value for $w$, the resource distribution was slightly unfavorable to the proposer, especially when selection was weak. This result is obtained when a deterministic rule is used for partner switching \cite{Miyaji2013} although the disadvantage was much weaker in our results.

The acceptance threshold of the responder decreased when $w$ was small, thereby supporting the intuitive prediction. Intuitively, larger opportunities for partner switching (higher $w$) were \textit{disadvantageous} to agents with a larger $q$ because rejecting the offer could lead to a 0 payoff and losing links with other players. However, the value of $q$ increased rapidly as $w$ increased. Furthermore, $p_A \ (p_B)$ and $q$ started to increase with almost the same value of $w$. This result suggested that the proposer's strategy is strongly related to the character of the network.

Note that if the acceptance by $both$ responders was required, the favoritism was not observed because there was no reason to distinguish two responders. Group fairness was similarly observed when $w$ was large enough. Our point is that co-evolutionary mechanism can lead to group fairness even if the game involves one strategically irrelevant player (see supplementary fig.S1 for this result). 

\begin{figure}[tbp]
\centering
\includegraphics[width = 55mm, trim= 0 15 0 0]{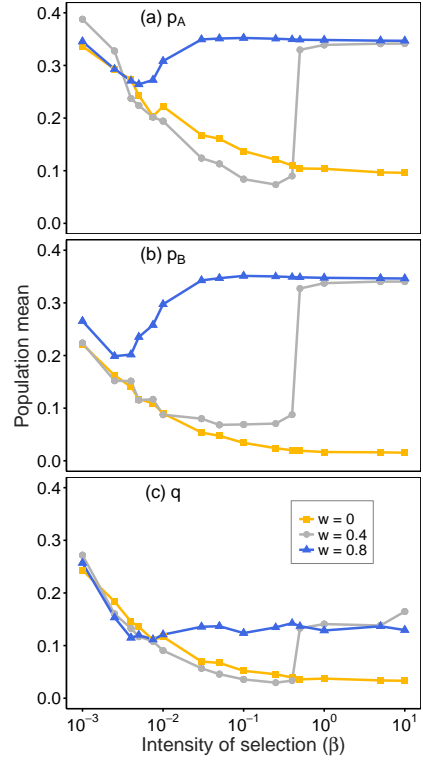}
\caption{\small Average offer ($p_A$, $p_B$) and acceptance threshold $q$ as a function of $\beta$ for different values of $w$. All the strategic variables were decreasing functions of $\beta$ without partner switching. A smaller value of $\beta$ was sufficient for the evolution of fairness if the frequency of partner switching was high. The responder's behavior ($q$) exhibited the same pattern with $p_A$ and $p_B$. Parameters: $N = 1000, \average{k} = 20$, and $\epsilon = 0.01$.}
\label{fig_papbq_beta}
\end{figure}

We also checked the results by investigating the strategy that emerged as a function of $\beta$. Figure \ref{fig_papbq_beta} shows that without the opportunity for partner switching ($w = 0$), all the strategic variables decreased as the intensity of selection became stronger. In fact, the offer to a similar agent ($p_A$) exceeded 0.3 when the intensity of selection was extremely weak, and this value was almost the same as the result in fig. \ref{fig_papbq_w} when the frequency of link adaptation was high. However, the value of $p_B$ (shown in panel (b)) was consistently smaller than the value of $p_A$ (panel (a)). Thus, a tendency toward favoritism was observed regardless of the value of $\beta$. The equality of the offer to the two agents was observed only after the combination of $w$ and $\beta$ was sufficiently large. Partner switching is beneficial to fair agents, so they can fully exploit this advantage under strong selection. We also observed that the value of $q$ exhibited a similar pattern to $p_A$ and $p_B$, which was similar to that shown in fig. \ref{fig_papbq_w}.

The effect of a higher frequency of partner switching can be understood in the following way. Agents who offer a small resource will lose edges and gain a smaller payoff. This fact encourages agents to make a generous offer (higher $p_A$ ($p_B$)). In addition, the smallest payoff determines the severance of relationships, so discriminating two responders is disadvantageous for the maintenance of edges. We noted that the offer size does not increase without limit. If the average of the sum of $p_A$ and $p_B$ is above 2/3, the worst payoff will probably be achieved when the focal agent is a proposer. Thus, the opportunity to sever relationships by unsatisfied responders will be reduced and a generous offer will not help agents to acquire more edges. In this situation, the acceptance threshold ($q$) is almost neutral unless it exceeds the offer size. An acceptance threshold above the offer size will simply lead to the rejection of a generous offer and the loss of social relationships. In fact, the value of $q$ fluctuated over time below $p_A$ ($p_B$) after the offer size became larger, which suggests that the acceptance threshold was not related to the resulting network when $w$ was large.

\begin{figure}[tbp]
\centering
\includegraphics[width = 55mm, trim= 0 15 0 0]{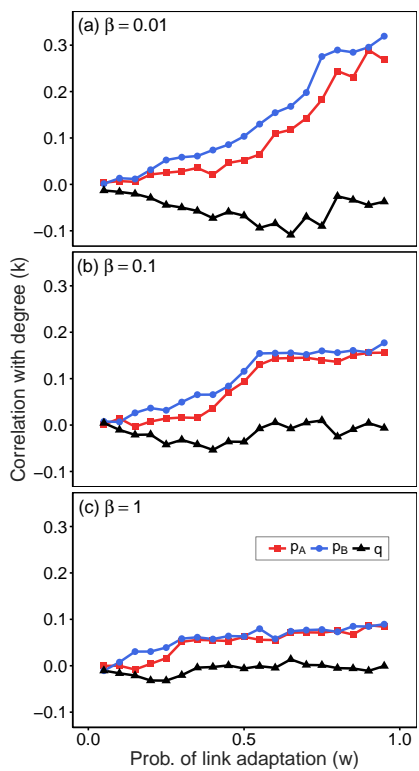}
\caption{\small Correlations between the strategic variables ($p_A$, $p_B$, $q$) and neighborhood size ($k$) as a function of $w$. The correlation between $p_A \ (p_B)$ and $k$ was positive and increased monotonically. By contrast, the correlation between $q$ and $k$ was negative with a smaller $w$ but approached zero with a larger $w$. Parameters: (a) $\beta = 0.01$, (b) $\beta = 0.1$, (c) $\beta = 1$; fixed $N = 1000, \average{k} = 20$, and $\epsilon = 0.01$.}
\label{fig_corpapbq_w}
\end{figure}

To examine the relationship between the game strategy and the resulting network, fig. \ref{fig_corpapbq_w} shows the Pearson's correlation coefficients between the strategy variables and the agent's neighborhood size ($k$) as a function of $w$ for different values of $\beta$. We computed the correlation in the last time period. The figure shows that the correlation between $p_A \ (p_B)$ and $k$ was positive and that it increased weakly as $w$ increased. This result suggests that the co-evolutionary mechanism penalizes greedy proposers by depriving them of opportunities for interaction. In addition, the correlation between $p_B$ and $k$ was stronger than the correlation between $p_A$ and $k$ when $w$ was small, which suggests that an offer to a disfavored agent was more important before a jump in the value of $p_A \ (p_B)$ was observed.

The figure also shows that the monotonic relationship did not hold with respect to $q$. With smaller values of $w$, the correlation between $q$ and $k$ was negative and it became stronger as the frequency of partner switching opportunities increased. The proposer kept their own share with modest values of $w$, and the higher demand threshold increased the possibility of bargaining breaking down (0 payoff) as well as risking the relationships with other players. However, with a larger value of $w$, the value of the correlation coefficient started to increase and it approached 0. Therefore, the relationship between the value of $q$ and the disadvantage due to the co-evolutionary mechanism disappeared. This pattern supports an interpretation where the co-evolutionary mechanism influences the behavior in the ultimatum game mainly through the proposer's strategy.

\begin{figure}[tbp]
\centering
\includegraphics[width = 55mm, trim= 0 15 0 0]{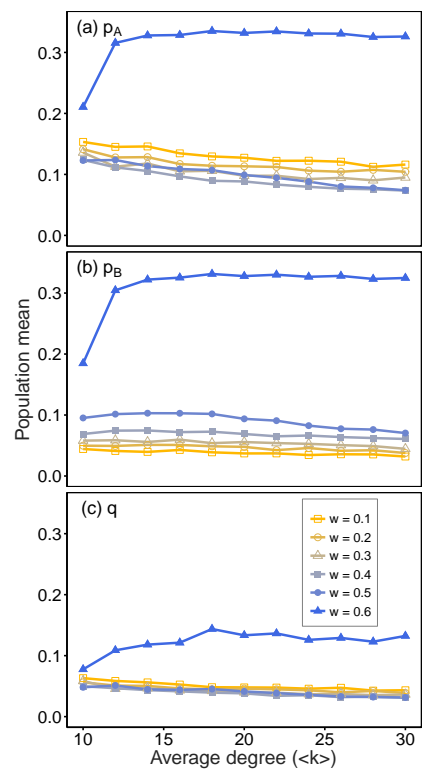}
\caption{\small Strategic variables as a function of \average{k}: (a) $p_A$, (b) $p_B$ and (c) $q$. We show the results obtained for different values of $w$. Basically, fairer behavior was observed as the average degree decreased. However, a small average degree prevented a fair strategy in some cases when the frequency of partner switching was high. Parameters: $N = 1000, \beta = 0.1$, and $\epsilon = 0.01$.}
\label{fig_papbq_k}
\end{figure}

\begin{figure}[tbp]
\centering
\includegraphics[width = 60mm, trim= 0 15 0 0]{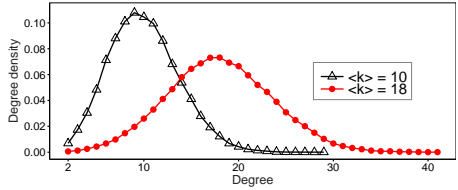}
\caption{\small Density of degree in the last period when $\average{k} = 10$ and $\average{k} = 18$. Larger average degree led to larger variance, which was beneficial for generous proposers when the value of $w$ (and the correlation between offer size and degree) was high. Parameters: $N = 1000, w = 0.6, \beta = 0.1$, and $\epsilon = 0.01$.}
\label{fig_density}
\end{figure}

Next, we examined the effect of the average degree on the emerging behavior. Figure \ref{fig_papbq_k} shows the results of TUG as the function of $\average{k}$ for different values of $w$. Basically, the results obtained in previous studies \cite{Page2000, Kuperman2008, Deng2011, Gao2011} were replicated and a smaller average degree led to fairness (higher $p_A$, $p_B$ and $q$; but see \cite{Wang2015d}). Because the number of successful bargaining rather than the gain from one game had a stronger effect on the responders ' payoff with larger \average{k}, responders had a stronger temptation to lower $q$ as the neighborhood size increased and the proposal also decreased accordingly \cite{Wang2014a, Wang2015d}. This logic also seemed to apply in our TUG. An exception to this pattern occurred when the value of $w$ became large and the variables approached the values observed in the fair state. For example, when $w = 0.6$, $p_A$ ($p_B$, $q$) exhibited a different pattern and a smaller $\average{k}$ hindered the fair strategy in some cases. Generous proposers enjoyed the benefit of larger degree when $w$ was high (fig.\ref{fig_corpapbq_w}). Because larger variance of degree was observed with larger \average{k} (fig. \ref{fig_density}, this result seems to be independent of the resulting strategy since the same pattern was observed when $w \leq 0.5$), smaller \average{k} lowered the benefit of degree heterogeneity for generous proposers. 

\begin{figure}[tbp]
\centering
\includegraphics[width = 55mm, trim= 0 15 0 0]{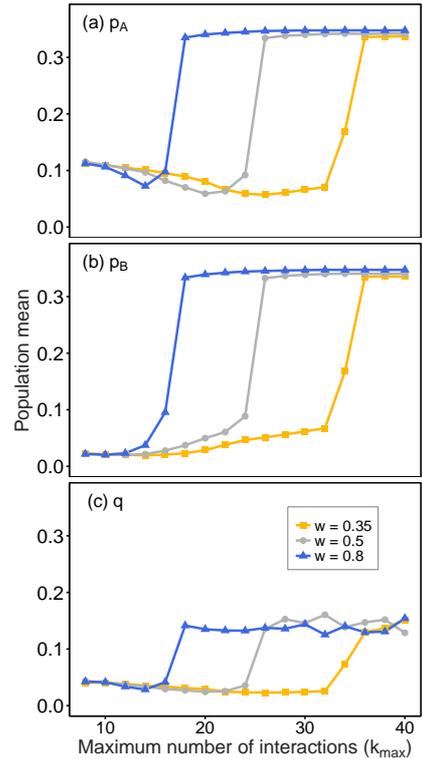}
\caption{\small Strategic variables as a function of the maximum number of interactions ($k_{max}$): (a) $p_A$, (b) $p_B$ and (c) $q$. We show the results obtained for different values of $w$. A smaller maximum number of interactions prevented the emergence of group fairness. A higher frequency of partner switching reduced the number of games required to achieve group fairness. Parameters: $N = 1000, \average{k} = 20, \beta = 1.0$, and $\epsilon = 0.01$.}
\label{fig_papbq_mk}
\end{figure}

Finally, we examined the impact of limiting the maximum number of games. The evolution of the network structure generated a heterogeneous neighborhood size (opportunities for social interactions) among agents. It is natural to assume that some agents engaged in more social interactions, but agents might not have been able to fully exploit the benefits of the larger neighborhood size. One method for dealing with this possibility is restricting the maximum number of interactions per unit time \cite{Poncela2011}. Thus, we restricted the number of games in terms of both strategy updating and partner switching. Figure \ref{fig_papbq_mk} shows the results of TUG as a function of the maximum number of interactions ($k_{max}$) for different values of $w$. As expected, limiting the number of social interactions had a disadvantageous effect on the emergence of group fairness. Less egalitarian results emerged using the combination of parameters where group fairness evolved in fig. \ref{fig_papbq_w}. However, with a higher frequency of partner switching ($w = 0.8$), agents did not have to fully exploit the full capacity of potential interactions and the number of games needed for fairness to emerge was less than the mean degree (in the figure, $\average{k} = 20$).

\section{Discussion}
In this study, we investigated the co-evolutionary TUG. In the TUG, there is a possibility that a proposer might discriminate in favor of a specific responder. Few studies have examined \textit{group} fairness despite its empirical significance. The impact of a co-evolutionary mechanism is ambiguous because it appears to foster fair proposals and a rational response at the same time, but our results showed that the opportunities for partner switching led to the emergence of group fairness. With a small opportunity for partner switching, the proposer favored one responder. By increasing the frequency of link adaptation, we then observed that equality among responders was achieved while proposers maintained their own large share. With a higher frequency of partner switching, group fairness was observed. This state emerged because the co-evolutionary mechanism worked mainly through the proposer's behavior. In fact, generous proposers achieved a larger neighborhood size and the responder's strategy had no impact on the resulting network. We also observed that a smaller average degree enhanced fair behavior although the opposite pattern was observed in some cases when the frequency of partner switching was high. Finally, a higher frequency of partner switching could support the evolution of fairness when the maximum number of games was restricted.

Our results have similarities with some observational studies. For example, anthropologists have reported that the social relationships of individuals who violated the norm were severed \cite{Guala2012}. Our simulation results suggest that severing relationships functions as punishment \cite{Fehr2003} and can actually support group fairness.

Our study helps understand the emergence of fairness beyond \textit{dyadic} relationships, but future extensions would be beneficial. First, to simplify the problem, we investigated the TUG with one proposer. Obviously, we could consider more complex games. For example, an $N$-person ultimatum game with one proposer \cite{Santos2015} would increase the complexity of the proposer's strategy. In addition, we could also consider a game with more than one proposer, which was investigated in an experimental study \cite{Roth1991}. Furthermore, evolutionary or imitation dynamics were assumed in the present study, but the robustness of the results should be examined using another learning rule \cite{Santos2016}. We consider that this line of research would further deepen our understanding of the evolution of fairness.


\begin{thebibliography}{10}
\expandafter\ifx\csname url\endcsname\relax\def\url#1{\texttt{#1}}\fi

\bibitem{Nowak2006}
\Name{Nowak M.~A.} \REVIEW{Science}{314}{2006}{1560}.

\bibitem{Szabo2007}
\Name{Szab{\'{o}} G. \and F{\'{a}}th G.} \REVIEW{Phys. Rep.}{446}{2007}{97}.

\bibitem{Tomasello2009}
\Name{Tomasello M.} \Book{{Why We Cooperate}} (MIT Press, Cambridge) 2009.

\bibitem{Guth1982}
\Name{G{\"{u}}th W., Schmittberger R. \and Schwarze B.} \REVIEW{J. Econ. Behav.
  Organ.}{3}{1982}{367}.

\bibitem{Camerer2003}
\Name{Camerer C.~F.} \Book{{Behavioral Game Theory: Experiments on Strategic
  Interaction}} (Princeton University Press, Princeton, NJ) 2003.

\bibitem{Fehr1999}
\Name{Fehr E. \and Schmidt K.~M.} \REVIEW{Q. J. Econ.}{114}{1999}{817}.

\bibitem{Andre2011a}
\Name{Andr{\'{e}} J.-B. \and Baumard N.} \REVIEW{J. Theor.
  Biol.}{289}{2011}{128}.

\bibitem{Santos2015a}
\Name{Santos F.~P., Santos F.~C. \and Paiva A.} \Book{{The Evolutionary Perks
  of Being Irrational}} in proc. of \Book{AAMAS 2015} (International Foundation
  for Autonomous Agents and Multiagent Systems, Istanbul, Turkey) 2015 pp.
  1847--1848.

\bibitem{Rand2013}
\Name{Rand D.~G., Tarnita C.~E., Ohtsuki H. \and Nowak M.~A.} \REVIEW{Proc.
  Natl. Acad. Sci.}{110}{2013}{2581}.

\bibitem{Nowak2000}
\Name{Nowak M.~A., Page K.~M. \and Sigmund K.}
  \REVIEW{Science}{289}{2000}{1773}.

\bibitem{Page2002}
\Name{Page K.~M. \and Nowak M.~A.} \REVIEW{Bull. Math. Biol.}{64}{2002}{1101}.

\bibitem{Page2000}
\Name{Page K.~M., Nowak M.~A. \and Sigmund K.} \REVIEW{Proc. R. Soc. B Biol.
  Sci.}{267}{2000}{2177}.

\bibitem{Kuperman2008}
\Name{Kuperman M.~N. \and Risau-Gusman S.} \REVIEW{Eur. Phys. J.
  B}{62}{2008}{233}.

\bibitem{Sinatra2009}
\Name{Sinatra R., Iranzo J., G{\'{o}}mez-Garde{\~{n}}es J., Flor{\'{i}}a L.~M.,
  Latora V. \and Moreno Y.} \REVIEW{J. Stat. Mech. Theory
  Exp.}{2009}{P09012}.

\bibitem{Szolnoki2012}
\Name{Szolnoki A., Perc M. \and Szab{\'{o}} G.} \REVIEW{Phys. Rev.
  Lett.}{109}{2012}{078701}.

\bibitem{Szolnoki2012a}
\Name{Szolnoki A., Perc M. \and Szab{\'{o}} G.} \REVIEW{EPL}{100}{2012}{28005}.

\bibitem{Wu2013}
\Name{Wu T., Fu F., Zhang Y. \and Wang L.} \REVIEW{Sci. Rep.}{3}{2013}{1550}.

\bibitem{Wang2014a}
\Name{Wang X., Chen X. \and Wang L.} \REVIEW{Sci. Rep.}{4}{2014}{4534}.

\bibitem{Chen2015a}
\Name{Chen W., Wu T., Li Z., Wu N. \and Wang L.} \REVIEW{EPL}{109}{2015}{68006}.
  
\bibitem{Wang2015d}
\Name{Wang X., Chen X. \and Wang L.} \REVIEW{J. Theor. Biol.}{380}{2015}{103}.

\bibitem{Duan2010}
\Name{Duan W.-q. \and Stanley H.~E.} \REVIEW{Phys. Rev. E}{81}{2010}{026104}.

\bibitem{Deng2011}
\Name{Deng L., Tang W. \and Zhang J.} \REVIEW{Physica A}{390}{2011}{4227}.

\bibitem{Gao2011}
\Name{Gao J., Li Z., Wu T. \and Wang L.} \REVIEW{EPL}{93}{2011}{48003}.

\bibitem{Miyaji2013}
\Name{Miyaji K., Wang Z., Tanimoto J., Hagishima A. \and Kokubo S.}
  \REVIEW{Chaos, Solitons Fractals}{56}{2013}{13}.

\bibitem{Yang2015}
\Name{Yang Z., Li Z., Wu T. \and Wang L.} \REVIEW{EPL}{109}{2015}{40013}.

\bibitem{Zimmermann2004}
\Name{Zimmermann M.~G., Egu{\'{i}}luz V.~M. \and {San Miguel} M.} \REVIEW{Phys.
  Rev. E}{69}{2004}{065102}.

\bibitem{Pacheco2006a}
\Name{Pacheco J.~M., Traulsen A. \and Nowak M.~A.} \REVIEW{Phys. Rev.
  Lett.}{97}{2006}{258103}.

\bibitem{Santos2006}
\Name{Santos F.~C., Pacheco J.~M. \and Lenaerts T.} \REVIEW{PLoS Comput.
  Biol.}{2}{2006}{e140}.

\bibitem{Fu2008}
\Name{Fu F., Hauert C., Nowak M.~A. \and Wang L.} \REVIEW{Phys. Rev.
  E}{78}{2008}{026117}.

\bibitem{Szolnoki2008b}
\Name{Szolnoki A., Perc M. \and Danku Z.} \REVIEW{EPL}{84}{2008}{50007}.

\bibitem{VanSegbroeck2008}
\Name{{Van Segbroeck} S., Santos F.~C., Now{\'{e}} A., Pacheco J.~M. \and
  Lenaerts T.} \REVIEW{BMC Evol. Biol.}{8}{2008}{287}.

\bibitem{Fu2009}
\Name{Fu F., Wu T. \and Wang L.} \REVIEW{Phys. Rev. E}{79}{2009}{036101}.

\bibitem{Szolnoki2009b}
\Name{Szolnoki A. \and Perc M.} \REVIEW{EPL}{86}{2009}{30007}.

\bibitem{Tanimoto2009}
\Name{Tanimoto J.} \REVIEW{Physica A}{388}{2009}{953}.

\bibitem{Perc2010}
\Name{Perc M. \and Szolnoki A.} \REVIEW{Biosystems}{99}{2010}{109}.

\bibitem{Cong2014}
\Name{Cong R., Wu T., Qiu Y.-Y. \and Wang L.} \REVIEW{Phys. Lett.
  A}{378}{2014}{950}.

\bibitem{Chen2016}
\Name{Chen W., Wu T., Li Z. \and Wang L.} \REVIEW{Physica A}{443}{2016}{192}.

\bibitem{Wu2009a}
\Name{Wu T., Fu F. \and Wang L.} \REVIEW{Phys. Rev. E}{80}{2009}{026121}.

\bibitem{Wu2009c}
\Name{Wu T., Fu F. \and Wang L.} \REVIEW{EPL}{88}{2009}{30011}.

\bibitem{Zhang2011}
\Name{Zhang C.~Y., Zhang J.~L., Xie G.~M. \and Wang L.} \REVIEW{Eur. Phys. J.
  B}{80}{2011}{217}.

\bibitem{Boehm2012}
\Name{Boehm C.} \REVIEW{Science}{336}{2012}{844}.

\bibitem{Hauert2002}
\Name{Hauert C., {De Monte} S., Hofbauer J. \and Sigmund K.}
  \REVIEW{Science}{296}{2002}{1129}.

\bibitem{Santos2008}
\Name{Santos F.~C., Santos M.~D. \and Pacheco J.~M.}
  \REVIEW{Nature}{454}{2008}{213}.

\bibitem{Szolnoki2009c}
\Name{Szolnoki A., Perc M. \and Szab{\'{o}} G.} \REVIEW{Phys. Rev.
  E}{80}{2009}{056109}.

\bibitem{Helbing2010}
\Name{Helbing D., Szolnoki A., Perc M. \and Szab{\'{o}} G.} \REVIEW{New J.
  Phys.}{12}{2010}{083005}.

\bibitem{Shi2010}
\Name{Shi D.-M., Zhuang Y. \and Wang B.-H.} \REVIEW{EPL}{90}{2010}{58003}.

\bibitem{Szolnoki2010a}
\Name{Szolnoki A. \and Perc M.} \REVIEW{EPL}{92}{2010}{38003}.

\bibitem{Xu2010a}
\Name{Xu Z., Wang Z., Song H. \and Zhang L.} \REVIEW{EPL}{90}{2010}{20001}.

\bibitem{Szolnoki2011b}
\Name{Szolnoki A., Szab{\'{o}} G. \and Perc M.} \REVIEW{Phys. Rev.
  E}{83}{2011}{036101}.

\bibitem{Chen2012}
\Name{Chen X., Szolnoki A., Perc M. \and Wang L.} \REVIEW{Phys. Rev.
  E}{85}{2012}{066133}.

\bibitem{Wu2014a}
\Name{Wu T., Fu F., Dou P. \and Wang L.} \REVIEW{Physica A}{413}{2014}{86}.

\bibitem{Chen2015}
\Name{Chen X., Szolnoki A. \and Perc M.} \REVIEW{Phys. Rev.
  E}{92}{2015}{012819}.

\bibitem{Chen2016a}
\Name{Chen M.-h., Wang L., Sun S.-w., Wang J. \and Xia C.-y.} \REVIEW{Phys.
  Lett. A}{380}{2016}{40}.

\bibitem{Santos2015}
\Name{Santos F.~P., Santos F.~C., Paiva A. \and Pacheco J.~M.} \REVIEW{J.
  Theor. Biol.}{378}{2015}{96}.

\bibitem{Santos2016}
\Name{Santos F.~P., Santos F.~C., Melo F.~S., Paiva A. \and Pacheco J.~M.}
  \Book{{Dynamics of Fairness in Groups of Autonomous Learning Agents}} in
  \Book{AAMAS 2016}, edited by \Name{Osman N. \and Sierra C.} Lecture Notes in
  Computer Science (Springer, Cham) 2016 p. 107--126.

\bibitem{Santos2005a}
\Name{Santos F.~C., Rodrigues J. \and Pacheco J.~M.} \REVIEW{Phys. Rev.
  E}{72}{2005}{056128}.

\bibitem{Traulsen2006}
\Name{Traulsen A., Nowak M.~A. \and Pacheco J.~M.} \REVIEW{Phys. Rev.
  E}{74}{2006}{011909}.

\bibitem{Fu2012}
\Name{Fu F., Tarnita C.~E., Christakis N.~A., Wang L., Rand D.~G. \and Nowak
  M.~A.} \REVIEW{Sci. Rep.}{2}{2012}{460}.

\bibitem{Poncela2011}
\Name{Poncela J., G{\'{o}}mez-Garde{\~{n}}es J. \and Moreno Y.} \REVIEW{Phys.
  Rev. E}{83}{2011}{057101}.

\bibitem{Guala2012}
\Name{Guala F.} \REVIEW{Behav. Brain Sci.}{35}{2012}{1}.

\bibitem{Fehr2003}
\Name{Fehr E. \and Fischbacher U.} \REVIEW{Nature}{425}{2003}{785}.

\bibitem{Roth1991}
\Name{Roth A.~E., Prasnikar V., Okuno-Fujiwara M. \and Zamir S.} \REVIEW{Am.
  Econ. Rev.}{81}{1991}{1068}.

\end{thebibliography}

\end{document}